\documentclass[preprint,10pt]{sigplanconf}
\usepackage[T1]{fontenc} %%%key to get copy and paste for the code
\usepackage[utf8]{inputenc} %%% to support copy and paste with accents
\usepackage{times}
\usepackage[scaled=0.85]{helvet}
\usepackage{graphicx}
\usepackage{ifthen}
\usepackage{xspace}
\usepackage{alltt}
\usepackage{latexsym}
\usepackage{url}            
\usepackage{amssymb}
\usepackage{amsfonts}
\usepackage{amsmath}
\usepackage{stmaryrd}
\usepackage{enumerate}
\usepackage[pdftex,colorlinks=true,pdfstartview=FitV,linkcolor=blue,citecolor=blue,urlcolor=blue]{hyperref}
\usepackage{xspace}
\usepackage{tikz}
\usepackage{fancybox}

\newboolean{showcomments}
\setboolean{showcomments}{false}
\ifthenelse{\boolean{showcomments}}
  {\newcommand{\bnote}[2]{
	\fbox{\bfseries\sffamily\scriptsize#1}
    {\sf\small$\blacktriangleright$\textit{#2}$\blacktriangleleft$}
   }
   
  }
  {\newcommand{\bnote}[2]{}
   
  }

\graphicspath{{figures/}}
\newcommand{\figref}[1]{Figure~\ref{fig:#1}}
\newcommand{\figlabel}[1]{\label{fig:#1}}

\newcommand{\commented}[1]{}
\newcommand{\secref}[1]{Section~\ref{sec:#1}}
\newcommand{\seclabel}[1]{\label{sec:#1}}

\newcommand{\eg}{\emph{e.g.,}\xspace}
\newcommand{\ie}{\emph{i.e.,}\xspace}

\newcommand{\enode}[1]{\tikz[remember picture]\node[fill,circle,scale=0.4] (#1) {};}  
\newcommand{\elink}[2]{\tikz[remember picture,overlay]\path[line width=1pt] (#1.center) edge (#2.center) {};}  
\newcommand{\efork}[2]{\tikz[remember picture,overlay]\path[line width=1pt,out=right,in=up,looseness=0.6] (#1.center) edge (#2.center) {};}  

\newcommand{\enewsession}{new session}

\newcommand{\ecreatemethod}{add}

\newcommand{\etag}[1]{\scalebox{0.7}{\sffamily\doublebox{#1}}}
\newcommand{\eheadt}[1]{\scalebox{0.8}{\sffamily\ovalbox{$\lhd$ #1}}}

\newcommand{\eheadof}[1]{\eheadt{h [#1]}}
\newcommand{\eheadbefore}{\eheadt{h \emph{(before)}}}
\newcommand{\eheadafter}{\eheadt{h \emph{(after)}}}
\newcommand{\cherrypicked}{\etag{redone}}

\newcommand{\eob}[1]{{\color{gray}#1}}
\newcommand{\etab}{\hspace{2mm}}

\newcounter{enodecounter}
\newcounter{elastnodecounter}

\newcommand{\efirstnode}{\setcounter{enodecounter}{1}\enode{1}}

\newcommand{\echildnodefrom}[1]{%
\stepcounter{enodecounter}%
\enode{\arabic{enodecounter}}%
\elink{#1}{\arabic{enodecounter}}}

\newcommand{\echildnode}{%
\setcounter{elastnodecounter}{\value{enodecounter}}
\echildnodefrom{\arabic{elastnodecounter}}}

\newcommand{\eforknodefrom}[1]{%
\stepcounter{enodecounter}%
\enode{\arabic{enodecounter}}%
\efork{#1}{\arabic{enodecounter}}}

\newcommand\esubcaption[1]{\begin{center}\emph{#1}\end{center}}

\newenvironment{eview}
    {\begin{alltt}\sffamily\small}
    {\end{alltt}}

\newenvironment{esession}
    {\begin{alltt}\sffamily\small}
    {\end{alltt}}

\newcommand\eitem{\enode{1} }
\newcommand\eitemn{{\color{lightgray}\enode{1}} }

\usepackage{url}            
\makeatletter
\def\url@leostyle{%
  \@ifundefined{selectfont}{\def\UrlFont{\sf}}{\def\UrlFont{\small\sffamily}}}
\makeatother
\begin{document}

\title{Representing Code History\\ with Development Environment Events}

\authorinfo{Mart\'in Dias \and Damien Cassou \and St\'ephane Ducasse}{
RMoD\\
Inria Lille--Nord Europe --- University of Lille --- Lifl}

\maketitle

\begin{abstract}
Modern development environments handle information about the intent of the programmer: for example, they use abstract syntax trees for providing high-level code manipulation such as refactorings; nevertheless, they do not keep track of this information in a way that would simplify code sharing and change understanding. In most Smalltalk systems, source code modifications are immediately registered in a transaction log often called a ChangeSet. Such mechanism has proven reliability, but it has several limitations. In this paper we analyse such limitations and describe scenarios and requirements for tracking fine-grained code history with a semantic representation. We present Epicea, an early prototype implementation. We want to enrich code sharing with extra information from the IDE, which will help understanding the intention of the changes and let a new generation of tools act in consequence.
\end{abstract}


%%%%%%% SECTION %%%%%%%
\section{Introduction}
\seclabel{Introduction}

Modern integrated development environments (IDEs) can have information about the intent of the programmer: they use abstract syntax trees (ASTs) and provide high-level code manipulation (such as refactorings \cite{Fowl99a}). Nevertheless, they do not keep track of this information in a way that would simplify code sharing and change understanding. For example, after a few hours of work, developers might want to separately share the different changes they have worked on: documentation improvements, bug fixes, and feature additions are better committed separately to facilitate review and backtracking. If each change were semantically recorded, making separate commits would be much simpler: for example, a method renamed could be seen as just one high-level operation instead of many lines removes and added.

In this paper we describe scenarios, requirements, and an early prototype, named Epicea,\footnote{\url{http://smalltalkhub.com/\#!/~MartinDias/Epicea}} for tracking code history with a semantic representation. Based on Epicea, we want to enrich code sharing with extra information from the IDE, which will help understanding the intention of the changes and let tools act in consequence. For example, when a library developer updates an API (\eg by renaming a method), he can provide a dedicated semantic change to the library users so that they can update their client code automatically. 

\paragraph{Structure of the paper.} In \secref{ChangeSets} we describe the problem in current Smalltalk systems. In \secref{Scenarios} a series of scenarios illustrate the key requirements for tracking changes semantically. We summarise such requirements in \secref{Requirements}. We present the design of our prototype in \secref{Epicea}. \secref{Screenshots} has screenshots of our prototype in action. After a short overview of related work in \secref{RelatedWork} we conclude in \secref{Conclusion}.

%%%%%%% SECTION %%%%%%%
\section{Analysis of Current Smalltalk Systems}
\seclabel{ChangeSets}

In most Smalltalk systems \cite{Gold89a} source code modifications are logged immediately after any editing operation in a transaction log, often called a ChangeSet.\footnote{\url{http://wiki.squeak.org/squeak/674}} This transaction log acts as a tape recording source code changes. The programmer can navigate different versions of the code without requiring a traditional version control system (VCS), such as git, svn and Monticello. In addition, if the execution of the system is interrupted (\eg the virtual machine crashes or the process is killed), then such a log can be explored to recover and replay the sequence of changes.

While this log mechanism has proven to be reliable over the years, it has the following problems:
\begin{description}

\item[Barely structured text.] There is a lack of abstraction. The log is a text file where each new event is appended at the end, as a sequence of chunks. Instead of representing the events in a declarative format, the events are written as executable commands. The idea is that by re-evaluating them the original change is reproduced. This format makes it difficult for tools to recover semantic information.

\item[Elementary model.] A ChangeSet records only class, package and method definitions. As a result, ChangeSet lacks information about class modifications or high-level events such as refactorings.

\item[Mixing sources and system events.] ChangeSets mix source management (the state of a system) with system event recording (the steps to go from one state to the next). The same model and format is used for ChangeSets and the traditional in Smalltalk fileIn/fileOut mechanism. As a result, not all the events can be recorded (\eg refactorings, package loading). In addition, the granularity of the events is often too coarse, leading to problems on recovery. For example, instance variable addition and class addition are indistinguishable. 

\item[Losing intermediate states.] ChangeSets only keeps track of what entities (\eg a class or method) has been modified. The intermediate states of such entities cannot be recovered but just the current one.
\end{description}

In this paper we introduce the notions of \emph{Log} and \emph{View} to fix the above-mentioned problems.

%%%%%%% SECTION %%%%%%%
\section{Scenarios for Changes as Programming Activity Traces}
\seclabel{Scenarios}

In this section we present several scenarios that illustrate the use of logs and their interplay in the IDE. We first define the vocabulary used in the rest of this paper.

\begin{description}

\item[Image.] In a Smalltalk environment, an image is a snapshot of all the objects of the system, \ie a memory dump: this includes both the objects of the software under execution but also the classes and methods at the moment of the snapshot. An image acts as a cache with preloaded packages and initialised objects.

\item[Session.] An image can be launched, modified, and saved many times. We call each one of these periods a session.

\item[Operation.] We refer with this word to an action performed in a session. An operation can either have a duration in time (\eg an expression evaluation) or be a punctual fact (\eg a class addition). An operation can trigger other operations. In \figref{VocabularyExample}, the list in the top represents a session where the developer has done three operations: (1) he has loaded the version 1 of a package named P using a VCS; (2) he has undone the addition of the class A from package P; (3) he has added a new class named B to package P. The light grey bullets and the horizontal alignment of the elements represent triggering (undoing the addition of class A has triggered the removal of A).

\item[Event.] We define an event as a representation of an operation. Some events represent a modification in the source code; we refer to them as \emph{code changes}. Sometimes we say that an event triggered another event when the operation that the former event represents triggered the operation that the latter event represents.

\item[Log.] A log contains events recorded from the IDE. This includes, for example, class additions, method redefinitions, and refactorings. If the user does not save or if the system crashes, the log and the image will become desynchronised: \ie the log will contain information that is not in the image. 

\item[Code unit.] In this paper we call code unit to a package, class, trait or method.

\item[View.] The log can have an overwhelming amount of information recorded about the system. This makes it difficult to understand the changes in a particular code unit. To solve this problem we include the concept of \emph{view}. In \figref{VocabularyExample}, views for the class A and package P are shown. The history of A is simple: it was added and then removed. The view of P is more complex: first, the class A was added, then this change got undone, and finally the class B got added (creating an implicit branch in the view). Each view has a head, marked as \eheadof{X}, which represents the current state in the system for the code unit X. The current head will be the parent of the next change that affects this code unit and the head will be updated to point to this new change.

\item[Commit.] We call \emph{commit} a particular version of source code stored in a VCS. In \figref{VocabularyExample}, we mark the last change performed during the load of version 1 with the tag \etag{P version 1}.

\end{description}

% EXAMPLE %
\begin{figure}[!htp]
\begin{center}

\begin{minipage}[b]{0.5\columnwidth}
\begin{esession}
\eitem \enewsession
\eitem load package P version 1
\eitemn\etab add package P
\eitemn\etab add A \etag{P version 1} 
\eitem undo (add A)
\eitemn\etab remove A
\eitem add B 
\end{esession}
\esubcaption{Log}
\end{minipage}

\bigskip

\begin{minipage}[b]{0.5\columnwidth}
\begin{eview}
\efirstnode   add package P
\eob{  \eforknodefrom{1} add A} \etag{P version 1} 
\echildnodefrom{1}   add B \eheadof{P} 
\end{eview}
\esubcaption{View of package P}
\end{minipage}%
\begin{minipage}[b]{0.5\columnwidth}
\begin{eview}
\efirstnode add A
\echildnode remove A \eheadof{A}
\end{eview}
\esubcaption{View of class A}
\end{minipage}

\caption{Example.}
\figlabel{VocabularyExample}
\end{center}
\end{figure}

%%%%% SUBSECTION %
\subsection{Logs Transcend Sessions}

Since a code unit can be edited over multiple sessions, the history of a code unit transcend history of images. In this section we discuss some scenarios that crosscut sessions.

\paragraph{Tie the events of several sessions.} 

In \figref{PlugInOtherHistories} we show the history of the package P accumulated over three sessions. The view ignore session boundaries. 

% PLUG-IN OTHER HISTORIES %
\begin{figure}[!htp]
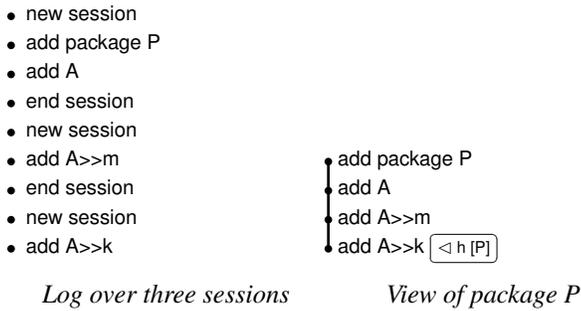

\begin{center}

\begin{minipage}[b]{0.5\columnwidth}
\begin{esession}
\eitem \enewsession
\eitem add package P 
\eitem add A
\eitem end session
\eitem \enewsession
\eitem add A>>m
\eitem end session
\eitem \enewsession
\eitem add A>>k
\end{esession}
\esubcaption{Log over three sessions}
\end{minipage}%
\begin{minipage}[b]{0.5\columnwidth}
\begin{eview}{}
\efirstnode add package P 
\echildnode add A
\echildnode add A>>m
\echildnode add A>>k \eheadof{P}
\end{eview}
\esubcaption{View of package P}
\end{minipage}
\caption{Views ignore session boundaries.}
\figlabel{PlugInOtherHistories}
\end{center}
\end{figure}

\paragraph{Recover lost changes after the IDE crashed.}
In \figref{Crash}, the user created a package P with a class A and committed the package P to a VCS. After adding  methods m and k, the IDE crashes. The user reopens the IDE, visualises the log of the crashed session, and redoes the lost changes. Such redone changes are shown as a new branch in the view. Each of those redone changes has a \cherrypicked~ tag. Such a tag always references the original entry so the developer can analyse the event in the context where it was originally logged.

% CRASH %
\begin{figure}[!htp]
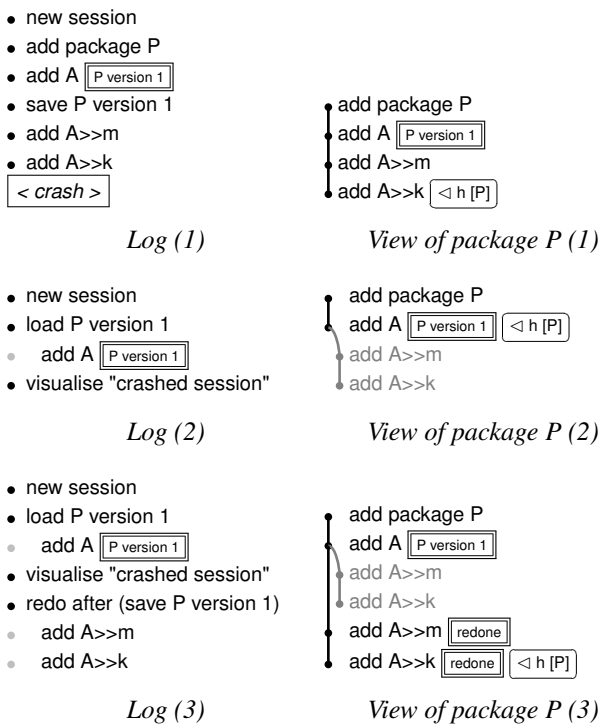

\begin{center}

%% ROW 1
\begin{minipage}[b]{\columnwidth}
\begin{minipage}[b]{0.5\columnwidth}
\begin{esession}
\eitem \enewsession
\eitem add package P
\eitem add A \etag{P version 1}
\eitem save P version 1
\eitem add A>>m
\eitem add A>>k
\fbox{\emph{< crash >}}
\end{esession}
\esubcaption{Log (1)}
\end{minipage}%
\begin{minipage}[b]{0.5\columnwidth}
\begin{eview}{}
\efirstnode add package P
\echildnode add A \etag{P version 1}
\echildnode add A>>m
\echildnode add A>>k \eheadof{P}
\end{eview}
\esubcaption{View of package P (1)}
\end{minipage}
\end{minipage}

\bigskip

%% ROW 2
\begin{minipage}[b]{\columnwidth}
\begin{minipage}[b]{0.5\columnwidth}
\begin{esession}
\eitem \enewsession 
\eitem load P version 1
\eitemn\etab add A \etag{P version 1}
\eitem visualise "crashed session" 
\end{esession}
\esubcaption{Log (2)}
\end{minipage}%
\begin{minipage}[b]{0.5\columnwidth}
\begin{eview}{}
\efirstnode   add package P
\echildnode   add A \etag{P version 1} \eheadof{P}
\eob{  \eforknodefrom{2} add A>>m
 \echildnode add A>>k}
\end{eview}
\esubcaption{View of package P (2)}
\end{minipage}
\end{minipage}

\bigskip

%% ROW 3
\begin{minipage}[b]{\columnwidth}
\begin{minipage}[b]{0.5\columnwidth}
\begin{esession}
\eitem \enewsession 
\eitem load P version 1
\eitemn\etab add A \etag{P version 1}
\eitem visualise "crashed session" 
\eitem redo after (save P version 1) 
\eitemn\etab add A>>m
\eitemn\etab add A>>k
\end{esession}
\esubcaption{Log (3)}
\end{minipage}%
\begin{minipage}[b]{0.5\columnwidth}
\begin{eview}{}
\efirstnode   add package P
\echildnode   add A \etag{P version 1}
\eob{  \eforknodefrom{2} add A>>m
 \echildnode add A>>k}
\echildnodefrom{2}   add A>>m \cherrypicked
\echildnode   add A>>k \cherrypicked \eheadof{P}
\end{eview}
\esubcaption{View of package P (3)}
\end{minipage}
\end{minipage}

\caption{Redo lost changes after the IDE crashed.}
\figlabel{Crash}
\end{center}
\end{figure}

\paragraph{Reload in fresh image.}
Since during experimentation images sometimes become unstable, it is a good practice to regularly rebuild from scratch the current head of development in a fresh image. Current infrastructure supports such practice by loading the code from the VCS, at the expense of losing the versions that occurred between two commits. The log overcomes such problems.

\subsection{Code Operations}

In this section we discuss some scenarios where navigation to previous versions of code or reorganisation of changes are important.

\paragraph{Undoing a code change.}
In \figref{Revert} we show that reverting the addition of method A>>m has different effects on the different views. In the package and class views, the original method additions are shown in grey as a branch. In that way, the original history of events with the original chronology is available to be browsed. In the A>>m view, the undo operation is seen as a removal of the method. For the A>>k view the operation has no impact. Note that in each view there is a head pointing to a different event.

% REVERT %
\begin{figure}[!htp]
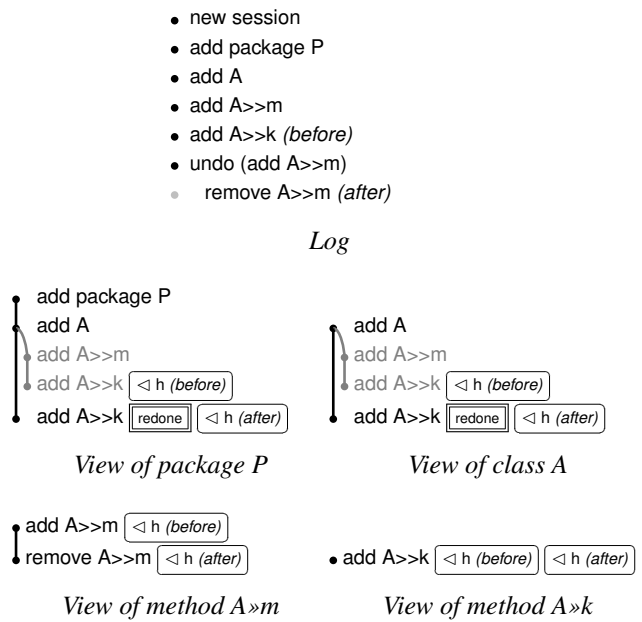

\begin{center}

\begin{minipage}[b]{0.5\columnwidth}
\begin{esession}
\eitem \enewsession
\eitem add package P 
\eitem add A
\eitem add A>>m
\eitem add A>>k \emph{(before)}
\eitem undo (add A>>m)
\eitemn\etab remove A>>m \emph{(after)}
\end{esession}
\esubcaption{Log}
\end{minipage}

\bigskip

\begin{minipage}[b]{0.5\columnwidth}
\begin{eview}
\efirstnode   add package P
\echildnode   add A
  \eob{\eforknodefrom{2} add A>>m
 \echildnode add A>>k} \eheadbefore
\echildnodefrom{2}   add A>>k \cherrypicked \eheadafter
\end{eview}
\esubcaption{View of package P}
\end{minipage}%
\begin{minipage}[b]{0.5\columnwidth}
\begin{eview}
\efirstnode   add A
  \eob{\eforknodefrom{1} add A>>m
 \echildnode add A>>k} \eheadbefore
\echildnodefrom{1}   add A>>k \cherrypicked \eheadafter
\end{eview}
\esubcaption{View of class A}
\end{minipage}

\bigskip

\begin{minipage}[b]{0.5\columnwidth}
\begin{eview}
\efirstnode add A>>m \eheadbefore
\echildnode remove A>>m \eheadafter
\end{eview}
\esubcaption{View of method A>>m}
\end{minipage}%
\begin{minipage}[b]{0.5\columnwidth}
\begin{eview}
\efirstnode add A>>k \eheadbefore \eheadafter
\end{eview}
\esubcaption{View of method A>>k}
\end{minipage}

\caption{Undoing the addition of A>>m. The operation has different effects at package, class and method level.} 
\figlabel{Revert}
\end{center}
\end{figure}

\paragraph{Grouping changes before committing.}
When a developer is working for some time on a project, chances are that he will perform multiple independent tasks. This happens even when there is a concrete goal such as implementing a new feature or fixing a bug: either a typo, or some code that deserves a refactoring, or any other change that is unrelated to the goal can appear. Tools should make it easy for a developer to fix the off-topic issue and let him either mark it or split it to a different branch so the main branch stays focused and cohesive. We need a kind of cherry picking of the elements we want to commit. In \figref{Split} we show an example of changes done in the package P, where the developer added a class B with some methods, and in the middle found and fixed a typo in the comment of A>>m. He decides to create a new branch to keep this change separated from the other ones. He also adds a comment to the separated change (modify A>>m) with a \etag{'typo fix'} tag.

% SPLIT %
\begin{figure}[!htp]
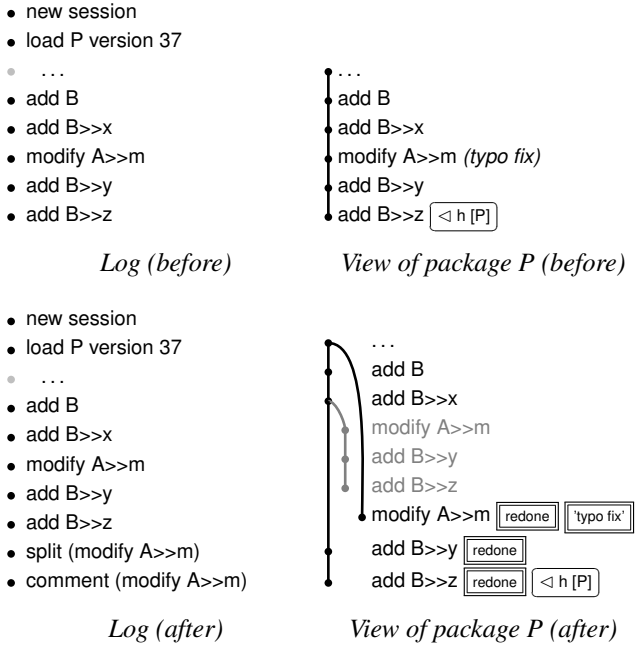

\begin{center}

\begin{minipage}[b]{0.5\columnwidth}
\begin{esession}
\eitem \enewsession
\eitem load P version 37
\eitemn\etab \dots
\eitem add B
\eitem add B>>x
\eitem modify A>>m
\eitem add B>>y 
\eitem add B>>z
\end{esession}
\esubcaption{Log (before)}
\end{minipage}%
\begin{minipage}[b]{0.5\columnwidth}
\begin{eview}{}
\efirstnode \dots
\echildnode add B
\echildnode add B>>x
\echildnode modify A>>m \emph{(typo fix)}
\echildnode add B>>y 
\echildnode add B>>z \eheadof{P}
\end{eview}
\esubcaption{View of package P (before)}
\end{minipage}

\bigskip

\begin{minipage}[b]{0.5\columnwidth}
\begin{esession}
\eitem \enewsession
\eitem load P version 37
\eitemn\etab \dots
\eitem add B
\eitem add B>>x
\eitem modify A>>m
\eitem add B>>y 
\eitem add B>>z
\eitem split (modify A>>m)
\eitem comment (modify A>>m)
\end{esession}
\esubcaption{Log (after)}
\end{minipage}%
\begin{minipage}[b]{0.5\columnwidth}
\begin{eview}{}
\efirstnode       \dots
\echildnode       add B
\echildnode       add B>>x
   \eob{\eforknodefrom{3}    modify A>>m
  \echildnode    add B>>y 
  \echildnode    add B>>z} 
      \eforknodefrom{1} modify A>>m \cherrypicked \etag{'typo fix'}
\echildnodefrom{3}       add B>>y \cherrypicked
\echildnode       add B>>z \cherrypicked \eheadof{P}
\end{eview}
\esubcaption{View of package P (after)}
\end{minipage}

\caption{Split changes for doing meaningful commits.}
\figlabel{Split}
\end{center}
\end{figure}

\paragraph{Commenting events.}
The developer can write arbitrary comments on an event (or group of events) to facilitate later understanding. We mentioned this feature in \figref{Split}, with the \etag{'typo fix'} tag. Additionally, the system can help the developer writing comments based on what triggered the related event.

\paragraph{Condensing code changes.}
The log might have changes that neutralise themselves (\eg a method is added and removed). In addition there are cases where the programmer may want to forget current history of certain entities. In \figref{Condense}, we show in an example how the condense operation works when applied to the package P. Without any optimisation, the operation is done in two main steps: first, undo the events until the older neutralised event (remove B, add C, and add B); second, redo only the needed changes (add C). 

% CONDENSE %
\begin{figure}[!htp]
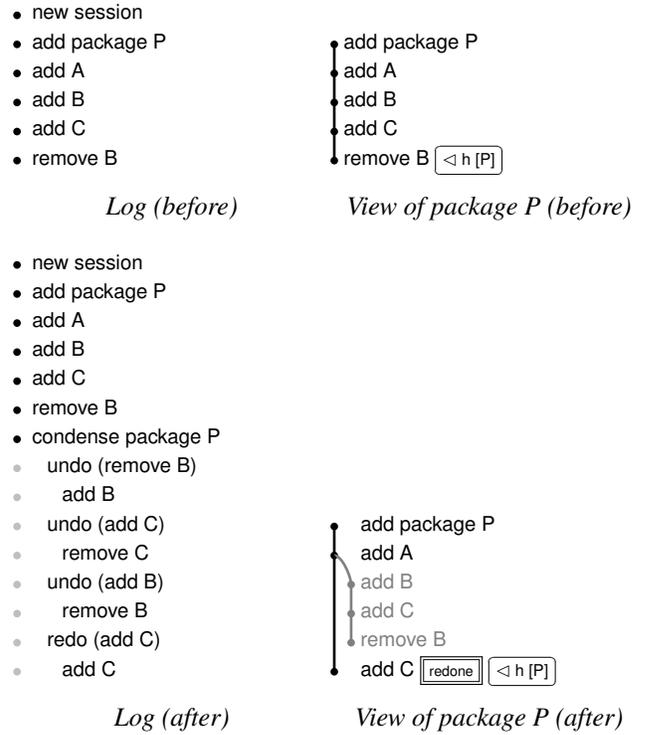

\begin{center}

\begin{minipage}[b]{0.5\columnwidth}
\begin{esession}
\eitem \enewsession
\eitem add package P
\eitem add A
\eitem add B
\eitem add C
\eitem remove B 
\end{esession}
\esubcaption{Log (before)}
\end{minipage}%
\begin{minipage}[b]{0.5\columnwidth}
\begin{eview}{}
\efirstnode add package P
\echildnode add A
\echildnode add B
\echildnode add C
\echildnode remove B \eheadof{P}
\end{eview}
\esubcaption{View of package P (before)}
\end{minipage}

\bigskip

\begin{minipage}[b]{0.5\columnwidth}
\begin{esession}
\eitem \enewsession
\eitem add package P
\eitem add A
\eitem add B
\eitem add C
\eitem remove B
\eitem condense package P
\eitemn\etab undo (remove B)
\eitemn\etab\etab add B 
\eitemn\etab undo (add C)
\eitemn\etab\etab remove C 
\eitemn\etab undo (add B)
\eitemn\etab\etab remove B 
\eitemn\etab redo (add C)
\eitemn\etab\etab add C 
\end{esession}
\esubcaption{Log (after)}
\end{minipage}%
\begin{minipage}[b]{0.5\columnwidth}
\begin{eview}{}
\efirstnode    add package P
\echildnode    add A
   \eob{\eforknodefrom{2} add B
  \echildnode add C
  \echildnode remove B}
\echildnodefrom{2}    add C \cherrypicked \eheadof{P}
\end{eview}
\esubcaption{View of package P (after)}
\end{minipage}

\caption{Condense operation.}
\figlabel{Condense}
\end{center}
\end{figure}

\paragraph{Recording refactoring information.}

Some high-level operations, such as refactorings, group events. In \figref{MethodRename}, a method is renamed (A>>m) and all senders (B>>k) of this method are updated. Each event related to the refactoring have a dedicated tag that references the high-level operation.

% METHOD RENAME %
\begin{figure}[!htp]
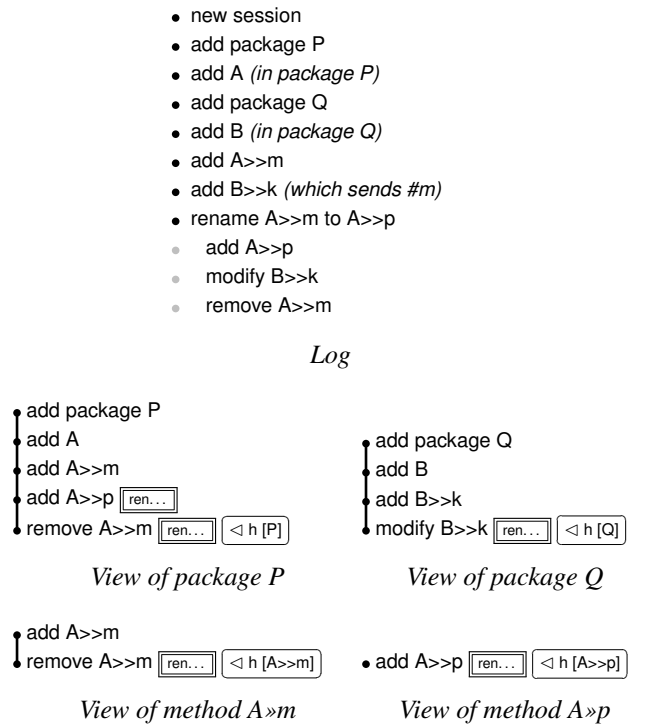

\begin{center}

\def\renmtop{ren\dots}

\begin{minipage}[b]{0.5\columnwidth}
\begin{esession}
\eitem \enewsession
\eitem add package P
\eitem add A \textit{(in package P)}
\eitem add package Q
\eitem add B \textit{(in package Q)}
\eitem add A>>m
\eitem add B>>k \textit{(which sends #m)}
\eitem rename A>>m to A>>p
\eitemn\etab add A>>p
\eitemn\etab modify B>>k
\eitemn\etab remove A>>m 
\end{esession}
\esubcaption{Log}
\end{minipage}

\bigskip

\begin{minipage}[b]{0.55\columnwidth}
\begin{eview}{}
\efirstnode add package P
\echildnode add A
\echildnode add A>>m
\echildnode add A>>p \etag{\renmtop}
\echildnode remove A>>m \etag{\renmtop} \eheadof{P}
\end{eview}
\esubcaption{View of package P}
\end{minipage}%
\begin{minipage}[b]{0.45\columnwidth}
\begin{eview}{}
\efirstnode add package Q
\echildnode add B
\echildnode add B>>k
\echildnode modify B>>k \etag{\renmtop} \eheadof{Q}
\end{eview}
\esubcaption{View of package Q}
\end{minipage}

\bigskip

\begin{minipage}[b]{0.55\columnwidth}
\begin{eview}{}
\efirstnode \ecreatemethod A>>m
\echildnode remove A>>m \etag{\renmtop} \eheadof{A>>m}
\end{eview}
\esubcaption{View of method A>>m}
\end{minipage}%
\begin{minipage}[b]{0.45\columnwidth}
\begin{eview}{}
\efirstnode \ecreatemethod A>>p \etag{\renmtop} \eheadof{A>>p}
\end{eview}
\esubcaption{View of method A>>p}
\end{minipage}%

\caption{Rename A>>m to A>>p. The method B>>k uses it so it is modified by the refactoring.}
\figlabel{MethodRename}
\end{center}
\end{figure}

\subsection{Sharing Events}

Logs and events can be shared between developers, projects, and images. 

\paragraph{Replaying a concrete event.}
When two projects are forks from each other, events of one fork can be replayed in the other.

\paragraph{Replaying the intent of a refactoring.}
When a library developer updates an API (\eg by renaming a method), he can provide high-level events which can be replayed by library users so that they can update their client code automatically.

%%%%%%% SECTION %%%%%%%
\section{Scenarios: an Analysis}
\seclabel{Requirements}

We analysed several existing code change representations: ChangeSets, RingC \cite{Uqui12b}, Cheops \cite{Ebra08a}, NewChangeSystem,\footnote{\url{http://smalltalkhub.com/\#!/~EzequielLamonica/NewChangeSystem}} and DeltaStreams.\footnote{\url{http://wiki.squeak.org/squeak/6001}}
From previous work and the scenarios presented above we define the following requirements.

\subsection{Requirements}

\begin{enumerate}
\item Replay and undo operations. Starting from the same or similar system, the information in the log should be enough for reconstructing the state of the system at any point of the log.

\item Log must be immediately persisted out of the volatile memory so information survives IDE crashes.

\item Log entries can have tags, \ie meta-information. A tag can reference another entry. Tags can be added after the entry has been persisted.

\item Events should be represented as first-class entities. 

\item The change model should support modelling many different types of changes: structural elementary changes (method definitions), composed ones (refactorings), and system changes such as expression evaluation, redo, and branch creation.
\end{enumerate}

%%%%%%% SECTION %%%%%%%
\section{Epicea}
\seclabel{Epicea}

We implemented Epicea, an early prototype of the log and the event model. It was developed in Pharo \cite{Blac09a}. Epicea model started as a branch of NewChangeSystem project and then was deeply modified and extended.

\subsection{Event Model}

\begin{figure}[!htp]
\begin{center}
\includegraphics[width=0.45\columnwidth]{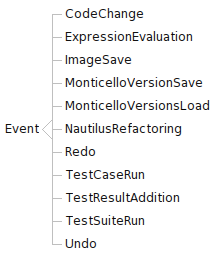}
\medskip
\includegraphics[width=\columnwidth]{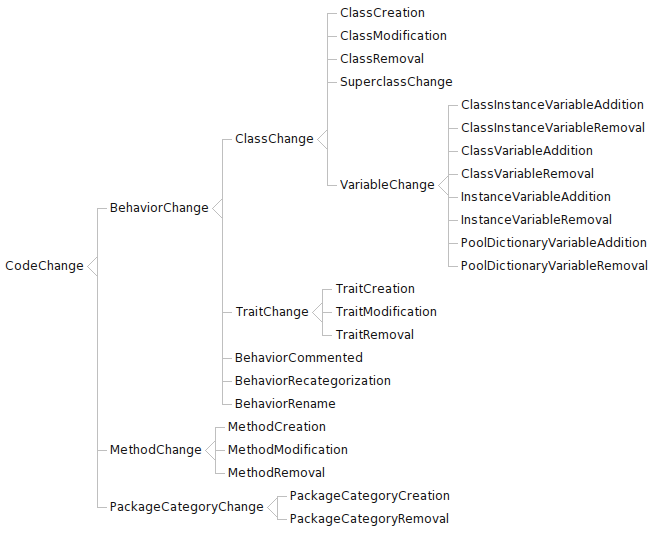}
\caption{The hierarchies of Event and CodeChange used in our prototype.}
\figlabel{ChangeHierarchy}
\end{center}
\end{figure}

In \figref{ChangeHierarchy} we show the class hierarchy of \emph{events} we implemented in Epicea. The most important sub-hierarchy is the one of CodeChange, which represents the operation that made the code change, such as class creation, method modification, etc. Code changes hold enough information about the operation performed for either reverting the change or redoing it. Epicea uses Ring definitions to take snapshots of the involved code units.

We need to record information about the situation in which events are logged. That is the timestamp when it was done, the author who did it, the potential event that triggered it (for example, undoing a method addition triggers a method removal). This meta-information of the event is stored in log entries, as explained below.

\subsection{Log Model}

In \figref{TreeSample} an object diagram shows how a log is represented in the prototype. A log has a head pointing to the entry where the upcoming entry will be attached. Each entry points to a parent entry and the content event. In \figref{TreeClasses} we show the design we implemented for Epicea. An entry has a dictionary of tags that allows attaching meta-information (author and timestamp). In the case of an event that triggers other events, each of these events has a tag pointing to the triggering event.

\begin{figure}[!htp]
\begin{center}
\begin{minipage}[b]{0.3\columnwidth}
\begin{esession}
\eitem \dots
\eitem load P version 1
\eitemn\etab add package P 
\eitemn\etab add A 
\end{esession}
%\bigskip
\esubcaption{Log}
\end{minipage}%

\begin{minipage}[b]{0.85\columnwidth}
\includegraphics[width=\textwidth]{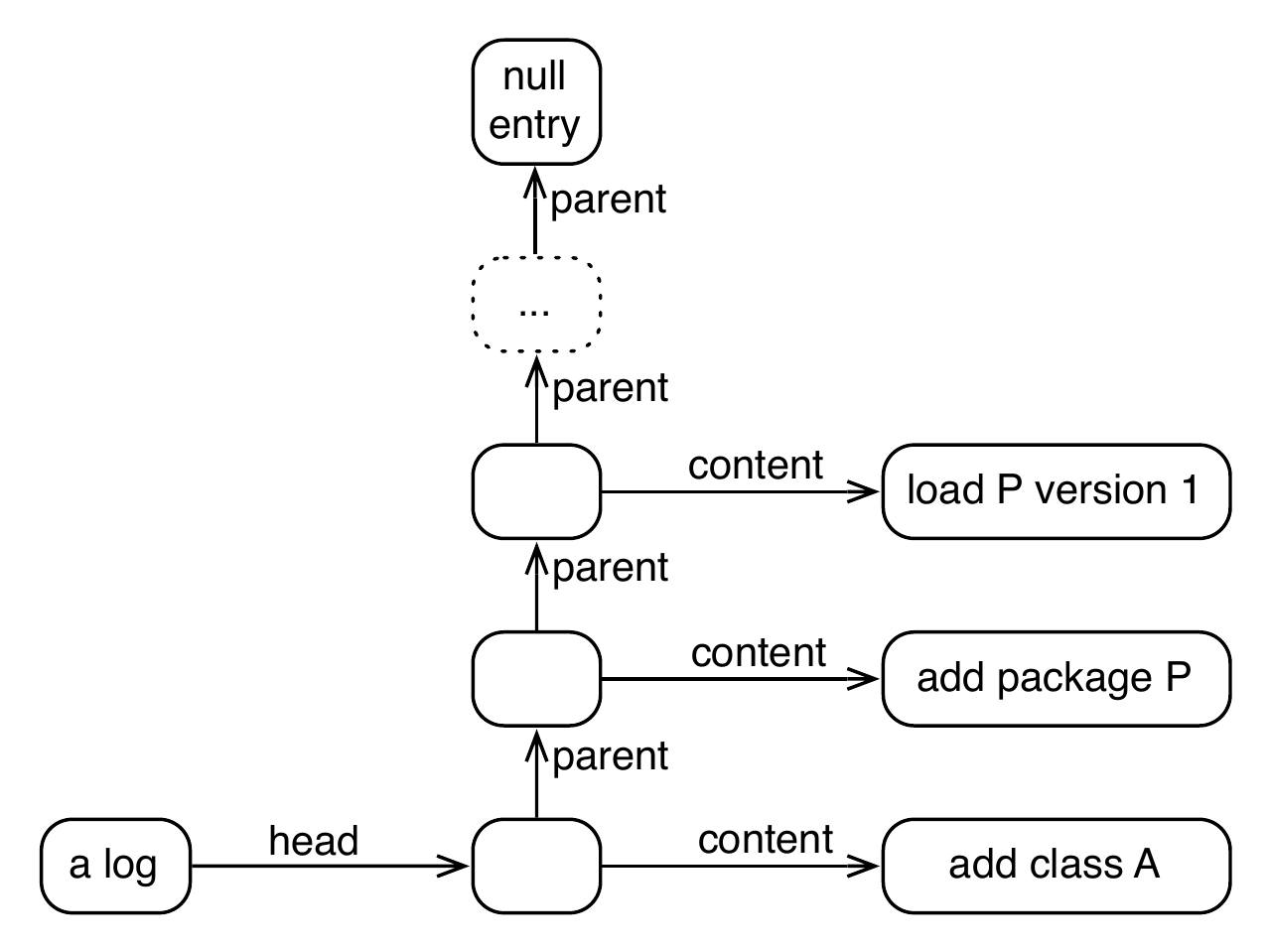}
\esubcaption{Internal representation}
\end{minipage}
\caption{Object diagram of an Epicea log.}
\figlabel{TreeSample}
\end{center}
\end{figure}

\begin{figure}[!htp]
\begin{center}
\includegraphics[width=0.9\columnwidth]{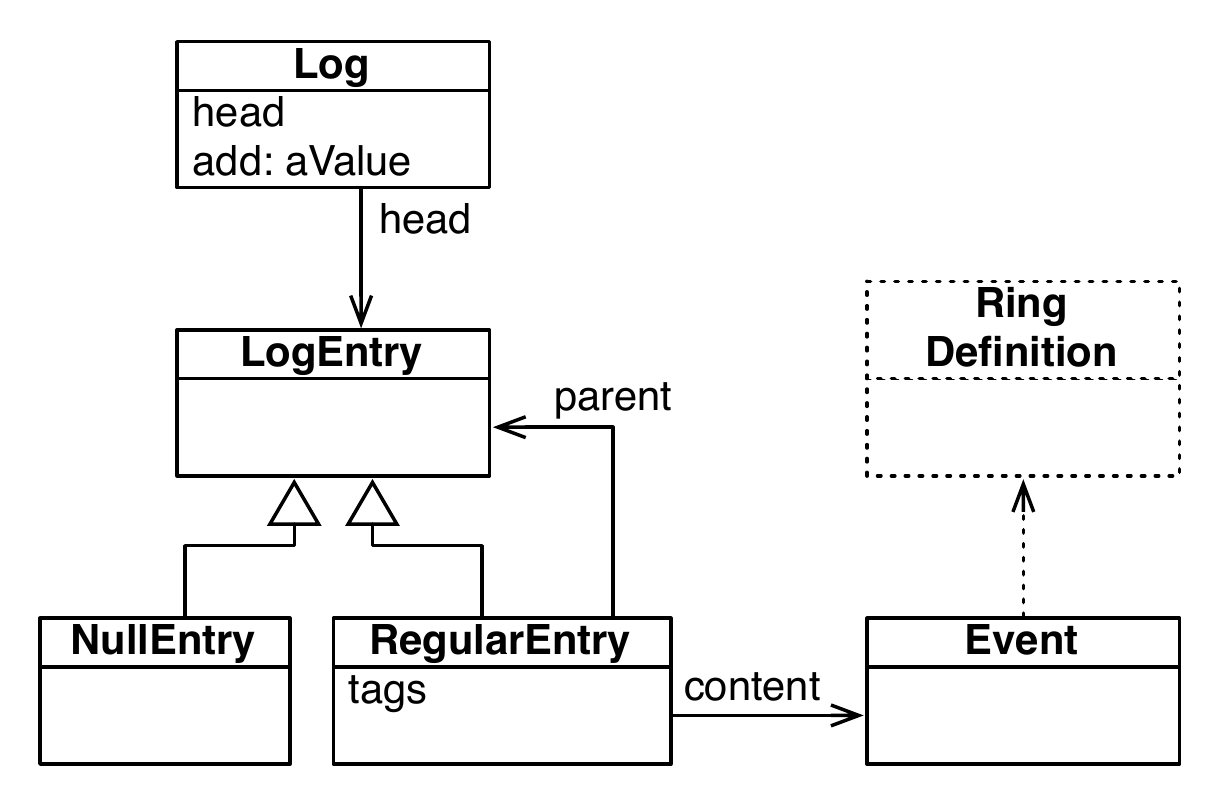}
\caption{Design of Epicea logs.}
\figlabel{TreeClasses}
\end{center}
\end{figure}

%%%%%%% SECTION %%%%%%%
\section{Revisiting the Scenarios}
\seclabel{Screenshots}

In \figref{ScreenshotLoad} an expression was evaluated. It triggered the load of the package named ConfigurationOfFuel. In turn, the load triggered many elemental code changes (package, class and method additions). In \figref{ScreenshotUndo} we show the log of an undo operation. The class A has been added in package P; then two methods have been added (A>>m and A>>k). Following, the undo of the addition of the method A>>m triggered the removal of such method. In \figref{ScreenshotRefactoring} we show a class rename refactoring as it is logged by Epicea.

\begin{figure}[!htp]
\begin{center}
\includegraphics[width=\columnwidth]{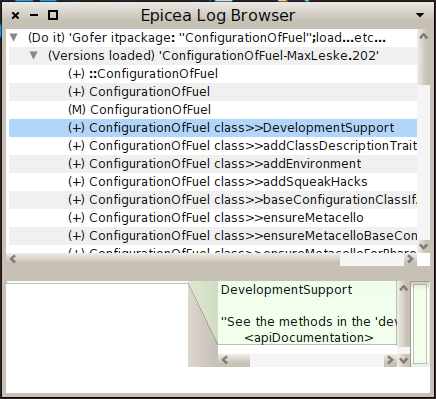}
\caption{Epicea log browser screenshot: an expression was evaluated. It triggered the load of the package named ConfigurationOfFuel. In turn, the load triggered many elemental code changes (package, class and method additions).}
\figlabel{ScreenshotLoad}
\end{center}
\end{figure}

\begin{figure}[!htp]
\begin{center}
\includegraphics[width=\columnwidth]{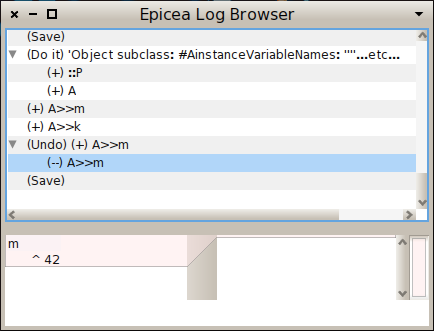}
\caption{Epicea log browser screenshot: Undo.}
\figlabel{ScreenshotUndo}
\end{center}
\end{figure}

\begin{figure}[!htp]
\begin{center}
\includegraphics[width=\columnwidth]{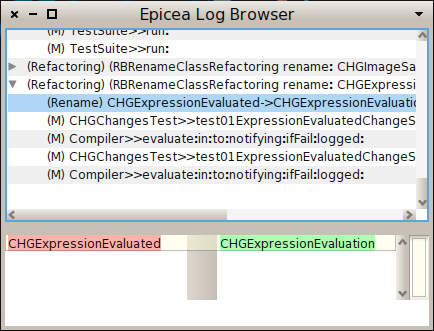}
\caption{Epicea log browser screenshot: Class rename refactoring.}
\figlabel{ScreenshotRefactoring}
\end{center}
\end{figure}

%%%%%%% SECTION %%%%%%%
\section{Related Work}
\seclabel{RelatedWork}

SpyWare \cite{Robb08b} captures and stores the code changes in a centralised repository in a extremely fine granularity. SpyWare records detailed changes such as a line added in a method, as well as more high-level changes like refactorings. The authors aim at post-mortem comprehension of developer work, while we focused on helping developers for their day-to-day work. 

CoExist \cite{Stein12b} is a Squeak/Smalltalk extension that preserves intermediate development states and provides immediate access to source code and run-time information of previous development states. CoExist allows for back-in time easily, automatic forks, inter-branch operations (such as rebase and cherry-pick). However, the authors do not talk of a persistence mechanism for the captured code changes. CoExist is not meant to be used to share code between images and projects. Still, CoExist is a source of inspiration for the Epicea model.
 
JET \cite{Uqui12b} allows analysing the dependencies between VCS versions. The authors extend the Ring meta-model \cite{Uqui11a} to perform the computations. Epicea uses Ring as well. It would be interesting to apply JET dependency analysis to logs to get fine-grained results.

%%%%%%% SECTION %%%%%%%
\section{Conclusion}
\seclabel{Conclusion}

Modern tools for sharing code lose extra information from IDE. We want to work on a new generation of tools that use such information to help understanding the intention behind code changes. In this paper we have presented our initial steps working in this direction. We have first described a series of scenarios that help discovering main requirements of our approach. Then, we have analyzed the problems found in current Smalltalk systems, focusing on the case of ChangeSets. Finally, we have presented our early prototype with an overview of the design, as well as some screenshots that show it in action.

%%%%%%% SECTION %%%%%%%

\section*{Acknowledgements} This work was supported by Ministry of Higher Education and Research, Nord-Pas de Calais Regional Council. 

\bibliographystyle{plain}
\bibliography{rmod,others}

\begin{thebibliography}{1}

\bibitem{Blac09a}
Andrew~P. Black, St\'ephane Ducasse, Oscar Nierstrasz, Damien Pollet, Damien
  Cassou, and Marcus Denker.
\newblock {\em Pharo by Example}.
\newblock Square Bracket Associates, Kehrsatz, Switzerland, 2009.

\bibitem{Ebra08a}
Peter Ebraert.
\newblock First-class change objects for feature-oriented programming.
\newblock In {\em Proceedings of the 15th Working Conference on Reverse
  Engineering}, WCRE'08, pages 319--322. IEEE Computer Society, 2008.

\bibitem{Fowl99a}
Martin Fowler, Kent Beck, John Brant, William Opdyke, and Don Roberts.
\newblock {\em Refactoring: Improving the Design of Existing Code}.
\newblock Addison Wesley, 1999.
\newblock ordered but not received.

\bibitem{Gold89a}
Adele Goldberg and Dave Robson.
\newblock {\em Smalltalk-80: The Language}.
\newblock Addison Wesley, 1989.

\bibitem{Robb08b}
Romain Robbes and Michele Lanza.
\newblock Spy{W}are: a change-aware development toolset.
\newblock In {\em Proceedings of the 30th International Conference on Software
  Engineering}, ICSE'08, pages 847--850, New York, NY, USA, 2008. ACM.

\bibitem{Stein12b}
Bastian Steinert, Damien Cassou, and Robert Hirschfeld.
\newblock {CoExist}: Overcoming aversion to change - preserving immediate
  access to source code and run-time information of previous development
  states.
\newblock In {\em DLS'12: Proceedings of the 8th Dynamic Languages Symposium},
  DLS '12, pages 107--118, New York, NY, USA, 2012. ACM.

\bibitem{Uqui12b}
Ver\'{o}nica Uquillas~G\'{o}mez.
\newblock {\em Supporting Integration Activities in Object-Oriented
  Applications}.
\newblock PhD thesis, Vrije Universiteit Brussel - Belgium \& Universit\'{e}
  Lille 1 - France, October 2012.

\bibitem{Uqui11a}
Ver\'onica Uquillas~G\'omez, St\'ephane Ducasse, and Theo D'Hondt.
\newblock Ring: a unifying meta-model and infrastructure for {S}malltalk source
  code analysis tools.
\newblock {\em Journal of Computer Languages, Systems and Structures},
  38(1):44--60, April 2012.

\end{thebibliography}
\end{document}